\documentclass[final,3p,times]{elsarticle}

\usepackage{graphicx}

\usepackage{amssymb}
\usepackage{amsmath}
\usepackage{amsbsy}

\journal{Journal of Non-Newtonian Fluid Mechanics}

\begin{document}
\begin{frontmatter}


\title{Periodic Exponential Shear of Complex Fluids}
\author[label1]{Chirag Kalelkar}
\author[label2]{Gareth H. McKinley}
\address[label1]{kalelkar@gmail.com}
\address[label2]{gareth@mit.edu}
\address{Hatsopoulos Microfluids Laboratory, Department of Mechanical Engineering, Massachusetts Institute of Technology, Cambridge, MA 02139, United States}
\begin{abstract}
We define a class of flows with exponential kinematics termed Periodic Exponential Shear (PES) flow which involve periodic exponential stretching of fluid elements along with their rotation. We exhibit analytical and numerical results for PES flow by using the Oldroyd-B model for viscoelastic fluids. We calculate the growth in the shear and the normal stresses analytically as well as demonstrate that repeated application of the flow leads to stable oscillatory shear and normal stresses. We define a material function applicable to a periodic, unsteady shear flow and show numerically that this material function exhibits deformation-rate thickening behavior for viscoelastic fluids subject to PES flow. We demonstrate the feasibility of PES flow by presenting preliminary experimental results from a controlled-strain rate rheometer, using a Newtonian mineral oil.
\end{abstract}
\end{frontmatter}
\section{Introduction}
Viscoelastic fluids flowing through complex geometries are often subject to intermittent flow regimes in which the fluid elements are both sheared as well as extended. Examples include the flow of drilling muds\cite{Handbook}, which are pumped from the surface of an oil rig down the drill string, through the drill bit, and back to the surface via an annular gap. Fluid flowing through a porous rock formation has a tortuous path through the pore space, which results in periodic shearing and extension of the fluid. In biological systems, the flow of blood through arteries is a periodic mixed flow. In flow through a microfluidic cross-slot device\cite{Simon}, fluid elements aligned with the centerline are subject to pure elongation coupled with translation, however, fluid elements which are off the symmetry axis are subject to a rapid (transient) stretch and rotation. The rotating (and stretching) fluid elements eventually align with a streamline and are convected thereafter. The rheological response of complex fluids under combined oscillatory-extensional flow is poorly understood, but critically important from a processing point of view. 

Several earlier studies\cite{Sivashinsky, Zulle,Dealy,Larson,Neergaard,Demarquette,Venerus,Kwan,Graham} have focussed on so-called {\it exponential shear} of viscoelastic fluids. Sivashinsky {\it et al.}\cite{Sivashinsky} measured the shear stress response of polyisobutylene solutions on a parallel-plate rheometer with an imposed aperiodic strain-rate of the form $\dot{\gamma}_{yx}(t;A,B)=Ae^{Bt}$ ($A$ and $B$ are variable parameters and the dot indicates a derivative with respect to time). Z\"ulle {\it et al.}\cite{Zulle} compared the extensional and shear viscosities measured in the uniaxial extension and unsteady shear flow of a low-density polyethylene melt, with the extensional strain-rate kept equal in magnitude to the shear strain-rate. Dealy and co-workers\cite{Dealy,Larson} probed the shear stress response of branched polymer melts subject to an aperiodic exponential strain of the form 
$\gamma_{yx}(t;a,\gamma_0)=\gamma_0(e^{at}-1)$, with $\gamma_0$ the strain amplitude and $a>0$ the growth rate of the exponential. The authors used an elastic dumbbell model with a Hookean spring to show that the high strains associated with such flows lead to rapid stretching of the polymer molecules within the fluid. Doshi and Dealy\cite{Dealy} concluded that exponential shear has a strong tendency to cause elongation of the polymer chains within the flow; however, Neergaard {\it et al.}\cite{Neergaard} carried out experiments with a linear polystyrene melt and found that the measured shear stresses in exponential shear are bounded by shear stress values measured during step-strain rate experiments at comparable instantaneous strain-rates. Their theoretical studies with a reptation model indicated that exponential shearing flow does indeed stretch entangled, linear polymer chains, but the chains are stretched to an equal extent during the inception of steady shear. Demarquette and Dealy\cite{Demarquette} carried out exponential shear tests using polystyrene solutions on a sliding-plate rheometer equipped with both a transducer and an apparatus for measuring flow-induced birefringence, and measured the shear stress as well as the third-normal stress coefficient [$\Psi_3(t;a,\gamma_0)\equiv(\tau_{xx}(t)-\tau_{zz}(t))/\dot{\gamma}_{yx}^2(t)$] as a function of the time. Their measurements showed that the shear stress in exponential shear fell below the linear viscoelastic prediction.

Venerus\cite{Venerus} utilized an aperiodic strain of the form $\gamma_{yx}(t;a)=2 \sinh(at)$, and measured the first normal stress difference [$N_1(t;a)\equiv\tau_{xx}(t)-\tau_{yy}(t)$] on a shear rheometer with cone and plate geometry, using a low-density polyethylene melt. He compared experimental data from constant strain-rate planar elongational and exponential shear flows, and concluded that exponential shear is incapable of generating stresses comparable to planar elongational flows, due to the non-zero vorticity in the flow (see Section $2$ below). Kwan {\it et al.}\cite{Kwan} carried out Brownian dynamics simulations of freely jointed bead-rod and bead-spring chain models subject to an aperiodic strain-rate of the form $\dot{\gamma}_{yx}(t;a)=2a \cosh(at)$, and compared their results with simulations of planar and uniaxial extensional flows. Graham {\it et al.}\cite{Graham} simulated the ``pom-pom" molecular model for branched polymer melts under aperiodic strain-rate of the form $\dot{\gamma}_{yx}(t;a)=ae^{at}$ and compared their results with the experimental data due to Z\"ulle {\it et al.}\cite{Zulle} and Venerus\cite{Venerus}. Several authors\cite{Zulle,Dealy,Larson,Kwan,Jobling,Lodge} have defined and measured appropriate material functions for exponential shear, which we discuss below in Section $3.4$.

In this paper, we introduce a non-linear rheometric deformation that we term {\it Periodic Exponential Shear} (PES) flow, which may be used for the investigation of material response to flow fields in which rapid stretching and rotation occur simultaneously and periodically. We evaluate analytical and numerical results using the Oldroyd-B model for viscoelastic fluids and define an appropriate material function for PES flow, which shear-thickens ({\it i.e.}, the material function for the flow grows faster than the linear viscoelastic prediction, as a function of the shear-rate). We demonstrate the utility of such a flow by showing preliminary experimental results from a shear rheometer, using a Newtonian mineral oil.
\section{Definition}
\begin{figure}[ht]
\includegraphics[height=2.25in]{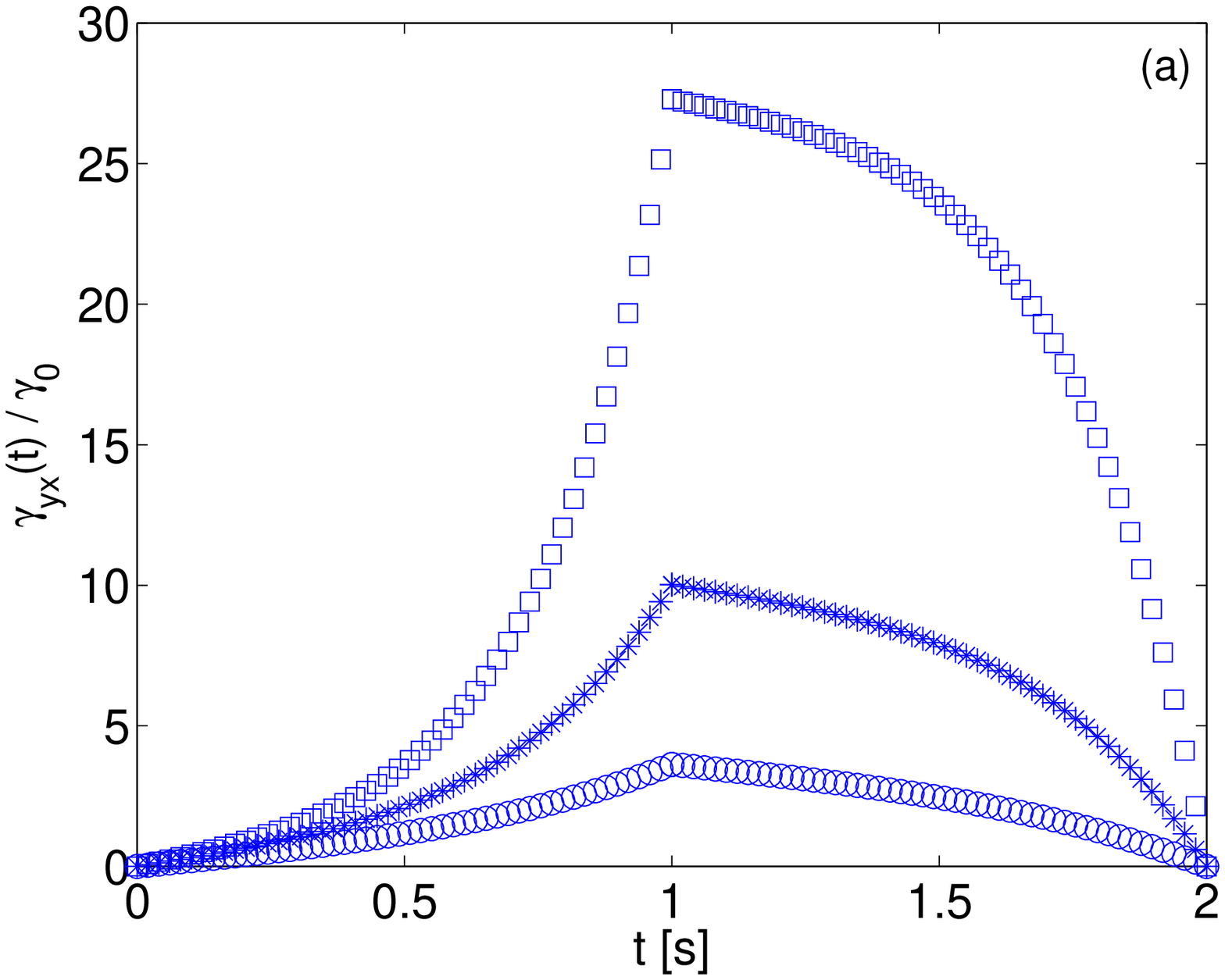}
\includegraphics[height=2.25in]{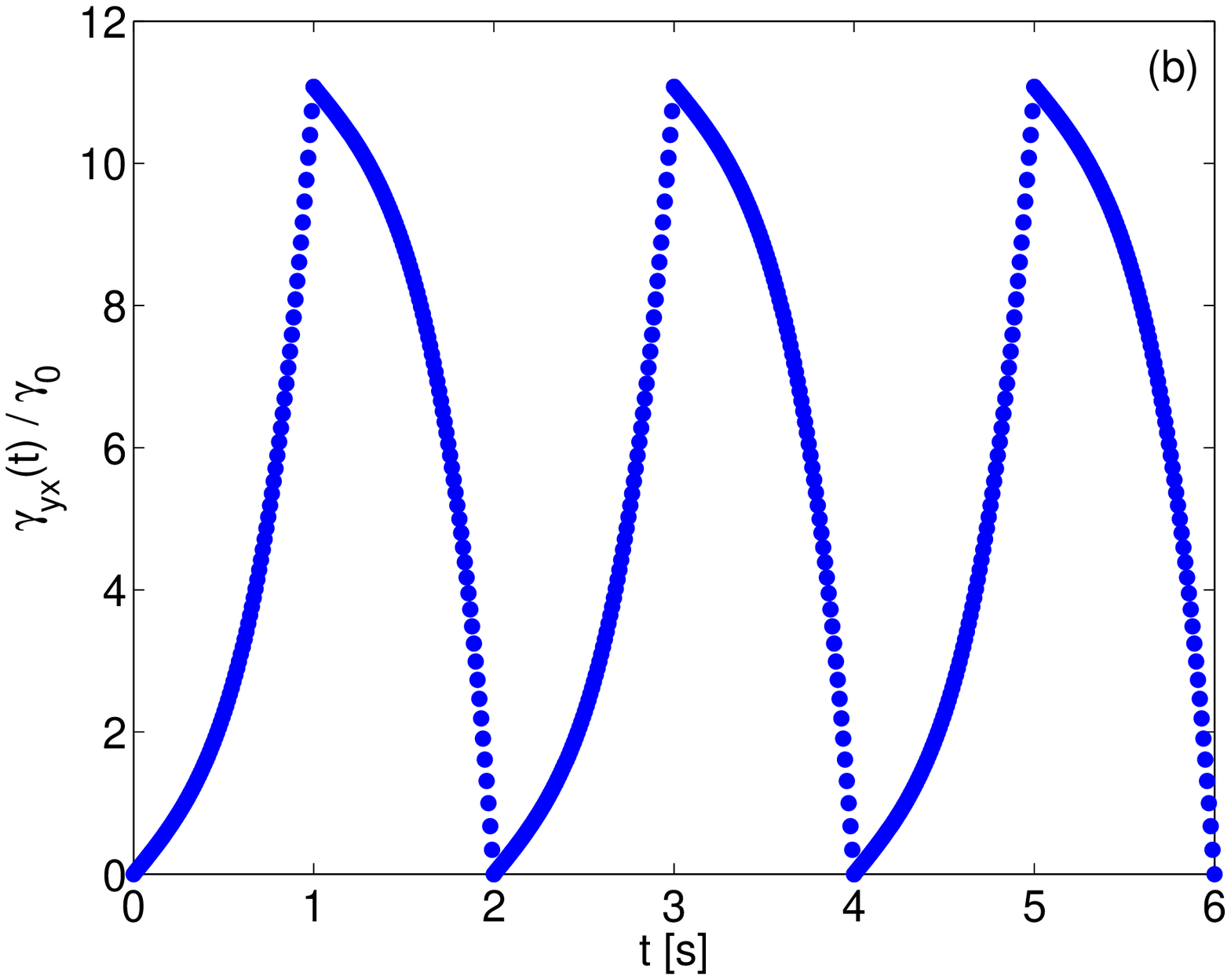}
\caption{(a) PES flow over one cycle of oscillation with $\gamma_0=1,t_0=0s,t_1=1s$ and varying exponential growth rates $a$: circle ($a=1s^{-1}$), star ($a=2s^{-1}$), square ($a=4s^{-1}$). (b) Periodic nature of PES flow demonstrated for $a=3.1s^{-1}, \gamma_0=1$.}
\label{fig:fig1}
\end{figure}

A general flow kinematics which incorporates periodic exponential stretching and rotation of the fluid elements, may be defined as:
\begin{align}\label{eq:PESgen}
\gamma_{yx}(t)=\gamma_0 \sinh[a(t-t_0)]\hspace{8.25cm}\textrm{for $t_0\le t\le t_1$},\nonumber\\
\gamma_{yx}(t)=\gamma_0 \sinh[a(t_1-t_0)]-\gamma_1 \sinh[b(t-t_1)]\hspace{1cm}\textrm{for $t_1\le t\le t_1+\frac{1}{b} \sinh^{-1}\left(\frac{\gamma_0}{\gamma_1} \sinh[a(t_1-t_0)]\right)$,}
\end{align}
with $b\ne0$, $\gamma_1\ne0$. Here $a$ and $b$ are the growth rates (units $s^{-1}$) of the exponentials, with strain amplitudes $\gamma_0$ and $\gamma_1$. The half-periods of this flow $\left[t_1-t_0,\frac{1}{b} \sinh^{-1}\left(\frac{\gamma_0}{\gamma_1} \sinh[a(t_1-t_0)]\right)\right]$ are unequal in general. For simplicity here, we choose identical growth rates $a=b$ and strain amplitudes $\gamma_0=\gamma_1$ in Eq. (\ref{eq:PESgen}), and define a periodic reversing (or PES) flow as:
\begin{eqnarray}
\gamma_{yx}(t)=\gamma_0 \sinh[a(t-t_0)]\hspace{4.5cm}\textrm{for $t_0\le t\le t_1$,}\nonumber\\
\gamma_{yx}(t)=\gamma_0 \sinh[a(t_1-t_0)]-\gamma_0 \sinh[a(t-t_1)]\hspace{1cm}\textrm{for $t_1\le t\le 2t_1-t_0$}.
\label{eq:PESeqns}
\end{eqnarray}
The half-periods of this derived flow are equal [$t_1-t_0$, $t_1-t_0$] and the time period for one cycle of deformation is $T=2(t_1-t_0)$. The parameters of PES flow are the exponential growth rate $a$, the strain amplitude $\gamma_0$ as well as the initial ($t_0$) and half-period time ($t_1$). In Fig. \ref{fig:fig1}(a), we plot the normalized strain as a function of the time for varying exponential growth rates $a$ over one cycle of oscillation with $\gamma_0=1,t_0=0s,t_1=1s$. In Fig. \ref{fig:fig1}(b), the periodic nature of PES flow is demonstrated for $a=3.1s^{-1}$, $\gamma_0=1$. The velocity field in an unsteady shear flow is given by ${\bf v}=(\dot{\gamma}_{yx}(t)y,0,0)$. PES flow is rotational, and the vorticity [$\boldsymbol{\omega}\equiv\triangledown\times{\bf v}=-\dot{\gamma}_{yx}(t){\bf\hat{z}}$] grows exponentially with time and oscillates. By contrast, planar elongational flow [${\bf v}=(ax,-ay,0)$] and uniaxial extensional flow [${\bf v}=(ax,-ay/2,-az/2)$] are irrotational, and fluid elements within such flows are subject to pure extension alone. We note that a frequently used flow-classification scheme (which does not take into account the vorticity of the flow) based on the invariants of the Finger strain tensor ${\bf C}^{-1}(t,t')$\cite{Venerus,Bird} yields the obvious result that PES may be classed as an unsteady shear flow and therefore the first and the second invariants of the tensor are equal ($I_1\equiv tr[{\bf C}^{-1}(t,t')]=I_2\equiv tr[{\bf C}(t,t')]$, here $tr$ indicates the trace of the tensor), as for the case of planar elongational flow.
The flow kinematics defined above ensure continuity of the strain over each cycle, but the strain-rate (and therefore the vorticity) is discontinuous in both sign and magnitude at the half-period time $t=t_1$. Our flow kinematics is experimentally realizable on a shear rheometer (see Section $3.5$ below), and directly comparable with the more traditional exponential shear flow in the first half-period.
\section{Results}
For our analytical and numerical studies, we use the well-known Oldroyd-B model\cite{Bird}, which serves as a useful representation for the flow of dilute polymeric liquids:
\begin{eqnarray}
\boldsymbol{\tau}+\lambda_1\boldsymbol{\tau}_{(1)}=\eta_0(\boldsymbol{\gamma}_{(1)}+\lambda_2\boldsymbol{\gamma}_{(2)}).
\end{eqnarray}
Here $\boldsymbol{\tau}$ is the stress tensor, $\boldsymbol{\gamma}$ is the strain tensor, $\eta_0$ is the zero-shear-rate viscosity, $\lambda_1$ and $\lambda_2$ are representative time-scales for the polymer and the solvent, respectively. The subscripts indicate the order of the upper-convected derivative, with 
$\boldsymbol{\sigma}_{(1)}\equiv\frac{\partial\boldsymbol{\sigma}}{\partial t}+{\bf v}\cdot\triangledown\boldsymbol{\sigma}-\{(\triangledown{\bf v})^T\cdot\boldsymbol{\sigma}+\boldsymbol{\sigma}\cdot(\triangledown{\bf v})\}$ for arbitrary operator $\boldsymbol{\sigma}$. The Oldroyd-B equations for an unsteady shearing flow are given by:
\begin{eqnarray}
\left(1+\lambda_1\frac{d}{dt}\right)\tau_{yx}(t)=\eta_0\left(1+\lambda_2\frac{d}{dt}\right)\dot{\gamma}_{yx}(t)
\label{eq:oldroydss}
\end{eqnarray}
\begin{eqnarray}
\left(1+\lambda_1\frac{d}{dt}\right)\tau_{xx}(t)=2\lambda_1\dot{\gamma}_{yx}(t)\tau_{yx}(t)+2\eta_0\lambda_2\dot
{\gamma}_{yx}^2(t)
\label{eq:oldroydns}
\end{eqnarray}
with $\tau_{yy}=\tau_{zz}=0$. The shear and the normal stress are assumed to be specified at some instant of time $t=t_0$, $\tau_{yx}(t_0)\equiv\tau_{yx}^0$ and $\tau_{xx}(t_0)\equiv\tau_{xx}^0$. As discussed below, the combination ${\cal G}\equiv\lambda_1a$ arises frequently in the solution to the above system of equations and we regard it as an effective dimensionless growth rate for PES flow. The numerical value of ${\cal G}$ does not determine the flow dynamics {\it uniquely}, since widely differing values of $\lambda_1$ and $a$ lead to different shear and normal stresses, but may still represent the same numerical value of ${\cal G}$. We note that the Oldroyd-B equations are non-singular for ${\cal G}=\{0.5,1\}$, but under PES flow, the shear stress $\tau_{yx}(t)$ is found to be singular for ${\cal G}=1$ and the normal stress $\tau_{xx}(t)$ is found to be singular for ${\cal G}=\{0.5,1\}$, as shown below. Hence, we solve for the different cases separately, in what follows. For our numerical work, we simulate Eqs. (\ref{eq:oldroydss}) and (\ref{eq:oldroydns}) using Matlab Ver. $7.9$ (The Mathworks Inc., United States).
\subsection{Shear stress $\tau_{yx}(t)$}
\begin{figure}[ht]
\includegraphics[height=2.25in]{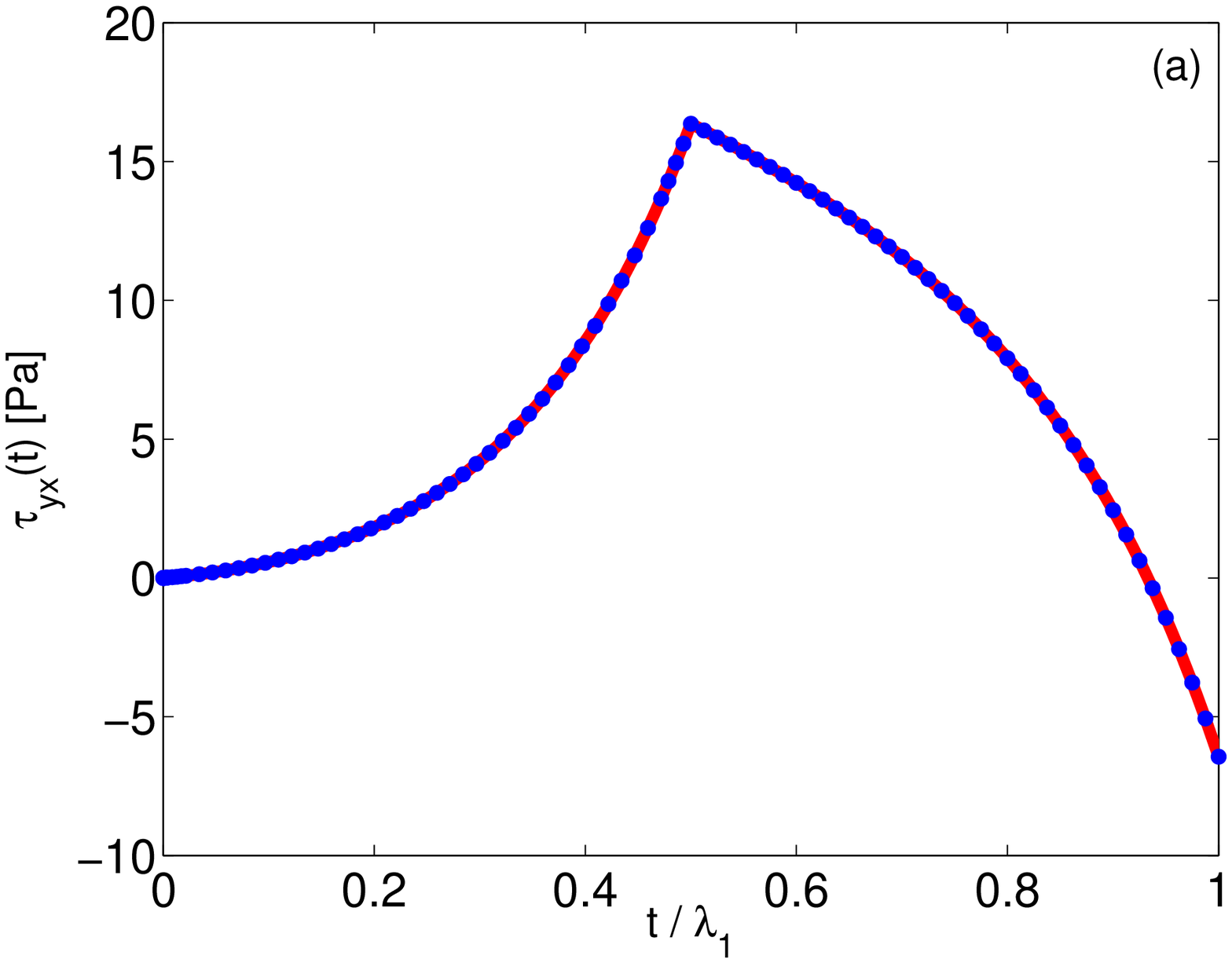}
\includegraphics[height=2.25in]{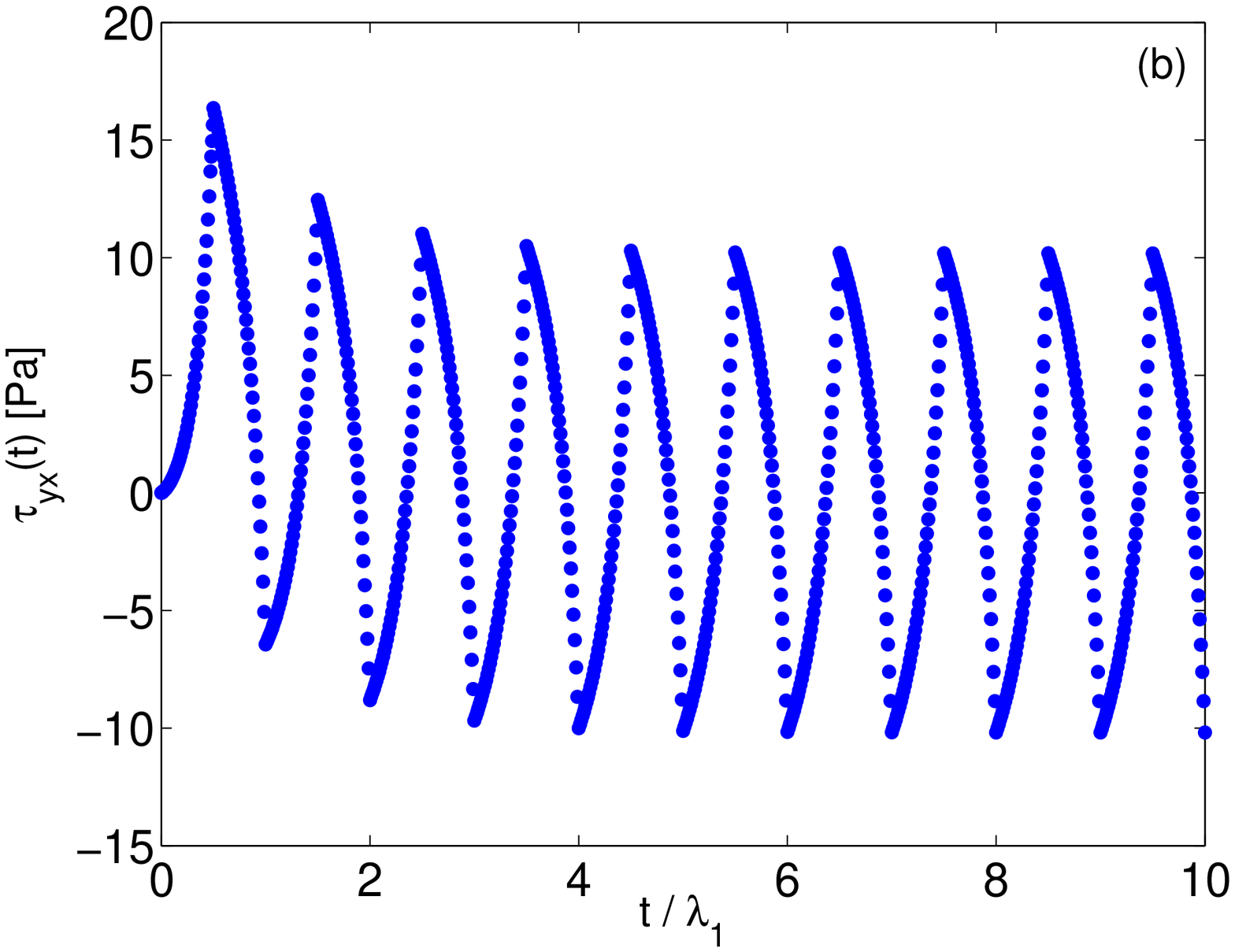}
\caption{(a) A plot of the analytical solution (red curve) for the shear stress $\tau_{yx}(t)$ as a function of the normalized time $t/\lambda_1$ and a comparison with the numerical solution (solid blue circles) of the Oldroyd-B model for $a=3s^{-1}, \gamma_0=1, \eta_0=1\hspace{0.1cm}\textrm{Pa.s.}, \lambda_1=2s, \lambda_2=1s$ (here ${\cal G}=6$). (b) A plot of the shear stress $\tau_{yx}(t)$ as a function of the normalized time $t/\lambda_1$ with the same parameters as above, for $10$ cycles of oscillation.}
\label{fig:ss_verify}
\end{figure}

\begin{figure}[ht]
\includegraphics[height=2.25in]{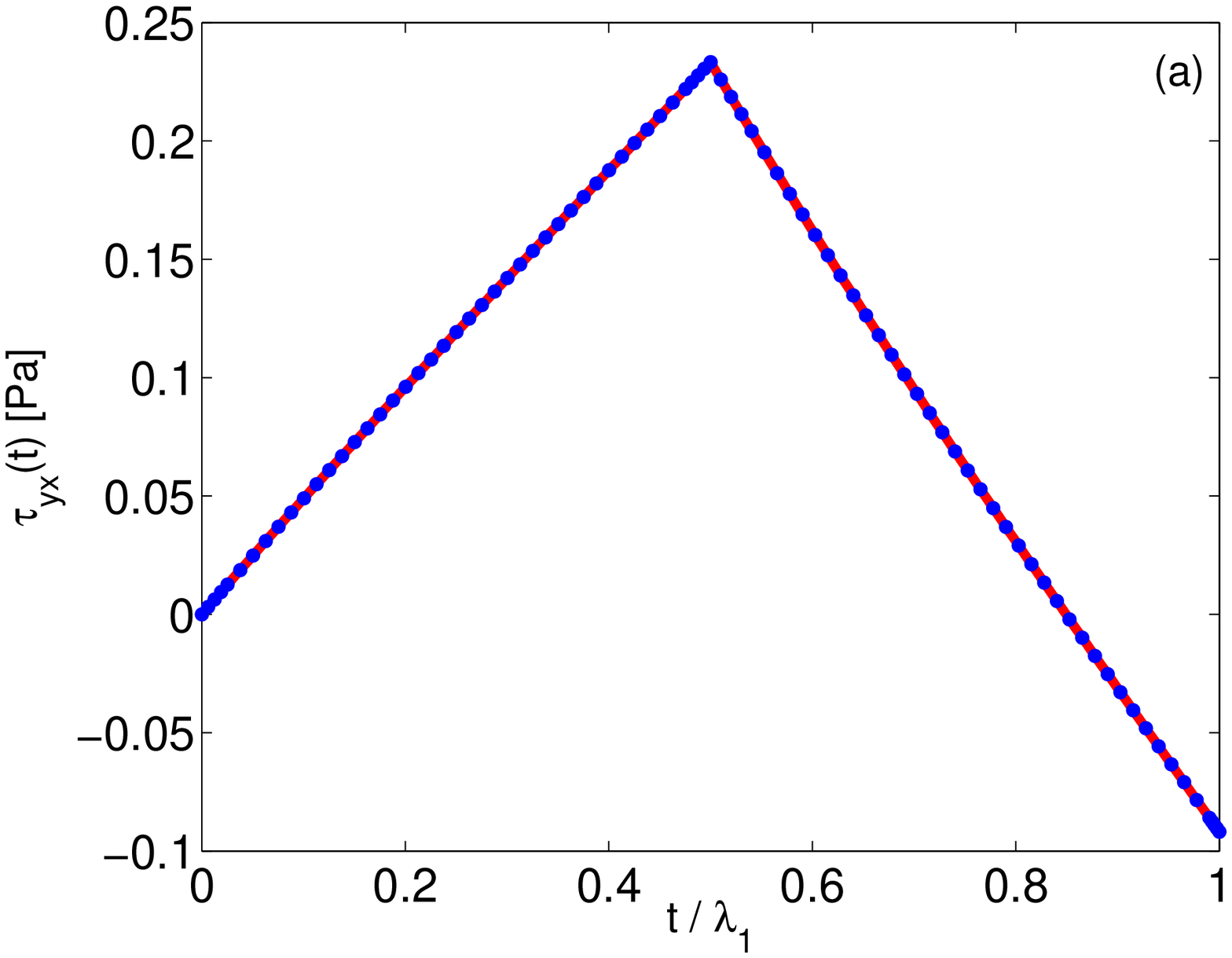}
\includegraphics[height=2.25in]{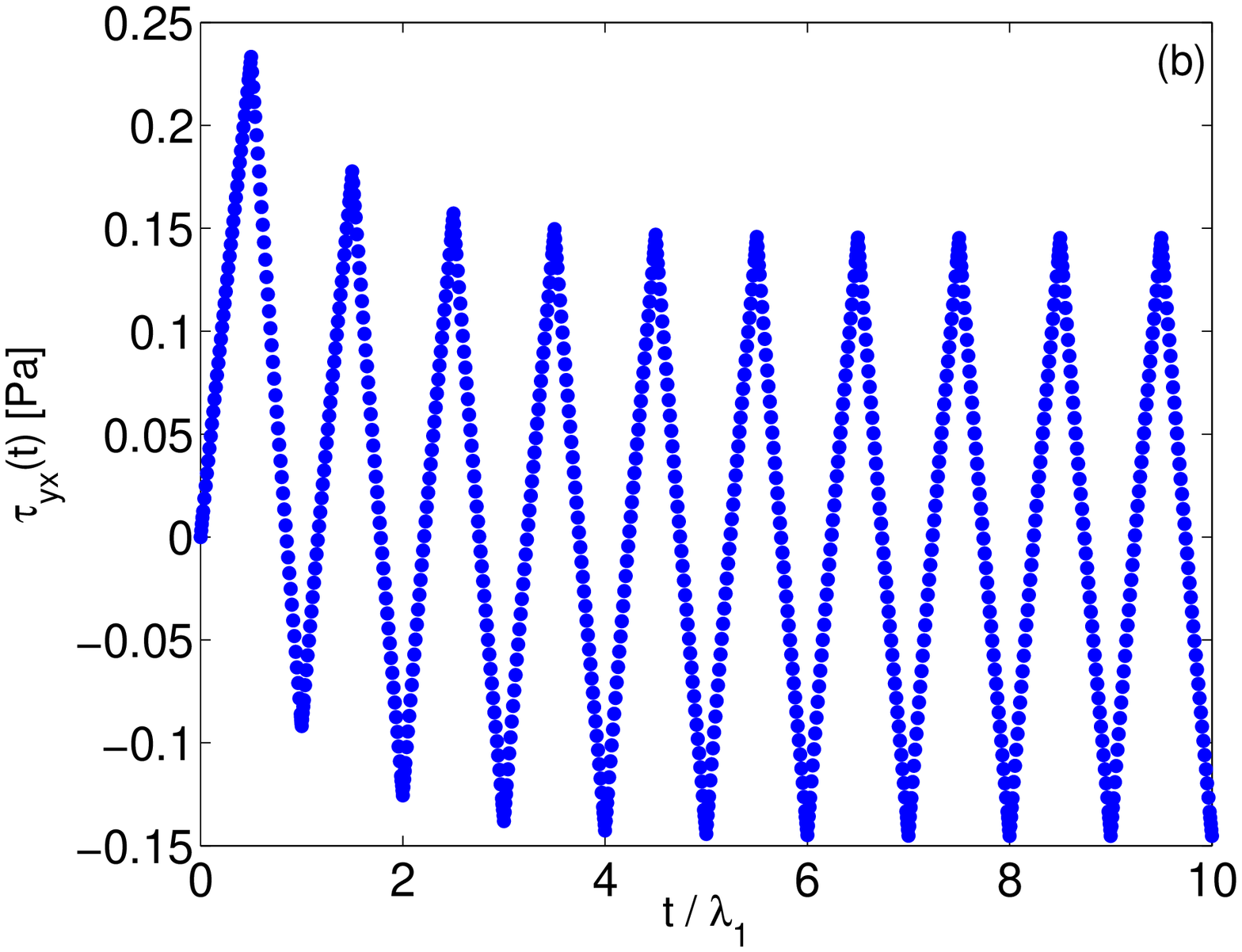}
\caption{(a) A plot of the analytical solution (red curve) for the shear stress $\tau_{yx}(t)$ as a function of the normalized time $t/\lambda_1$ and a comparison with the numerical solution (solid blue circles) of the Oldroyd-B model for $a=0.5s^{-1}$, $\gamma_0=1$, $\eta_0=1\hspace{0.1cm}\textrm{Pa.s.}$, $\lambda_1=2s$, 
$\lambda_2=1s$ (here ${\cal G}=1$). (b) A plot of the shear stress as a function of the normalized time with the same parameters as above, for 10 cycles of oscillation.}
\label{fig:qq}
\end{figure}

The solution of the linear first-order ordinary differential equation [Eq. (\ref{eq:oldroydss})] for $t_0\le t\le t_1$, ${\cal G}\ne1$ with the strain-rates calculated from the PES flow strain profile [Eq. (\ref{eq:PESeqns})] is:
\begin{eqnarray}\label{eq:oldss1}
\tau_{yx}(t)=\left[\tau_{yx}^0+\frac{\eta_0\gamma_0a(1-a^2\lambda_1\lambda_2)}{(\lambda_1a)^2-1}\right]
e^{-(t-t_0)/\lambda_1}+\frac{\eta_0\gamma_0a}{(\lambda_1a)^2-1}\{\lambda_1a \sinh(a(t-t_0))- \cosh(a(t-t_0))\}\\\nonumber
+\frac{\eta_0\gamma_0a^2\lambda_2}{(\lambda_1a)^2-1}\{\lambda_1a \cosh(a(t-t_0))- \sinh(a(t-t_0))\}.
\end{eqnarray}
The solution for $t_1\le t\le 2t_1-t_0$, ${\cal G}\ne1$ is:
\begin{eqnarray}\label{eq:oldss2}
\tau_{yx}(t)=\left[\tau_{yx}^1-\frac{\eta_0\gamma_0a(1-a^2\lambda_1\lambda_2)}{(\lambda_1a)^2-1}\right]e^{-(t-t_1)/\lambda_1}-\frac{\eta_0\gamma_0a}{(\lambda_1a)^2-1}\{\lambda_1a \sinh(a(t-t_1))- \cosh(a(t-t_1))\}\\\nonumber
-\frac{\eta_0\gamma_0a^2\lambda_2}{(\lambda_1a)^2-1}\{\lambda_1a \cosh(a(t-t_1))- \sinh(a(t-t_1))\}.
\end{eqnarray}
Here $\tau_{yx}^1$ equals the value of $\tau_{yx}(t_1)$ obtained from Eq. (\ref{eq:oldss1}), which ensures continuity of $\tau_{yx}(t)$ at $t=t_1$. In Fig. \ref{fig:ss_verify}(a), Eqs. (\ref{eq:oldss1}) and (\ref{eq:oldss2}) are plotted as red curves, and shown to agree with the numerical solution (solid blue circles) of the Oldroyd-B equations over one cycle of oscillation for the representative parameter values $a=3s^{-1}, \gamma_0=1, \eta_0=1\hspace{0.1cm}\textrm{Pa.s.}, \lambda_1=2s, \lambda_2=1s$.
In Fig. \ref{fig:ss_verify}(b), we plot the shear stress as a function of the normalized time for $10$ cycles of oscillation with the same set of parameter values. For ${\cal G}=1$, Eq. (\ref{eq:oldroydss}) is solved with $a=1/\lambda_1$, $\lambda_1\ne0$ using the strain-rates calculated from Eq. (\ref{eq:PESeqns}). The solution for $t_0\le t\le t_1$ is:
\begin{align}\label{eq:q1stress1}
\tau_{yx}(t)=\left[\tau_{yx}^0-\frac{\eta_0\gamma_0a(1+a\lambda_2)}{4}\right]e^{-a(t-t_0)}+\frac{\eta_0\gamma_0a^2(1-a\lambda_2)}{2}(t-t_0)e^{-a(t-t_0)}+\frac{\eta_0\gamma_0a(1+a\lambda_2)}{4}e^{a(t-t_0)}.
\end{align}
The solution for $t_1\le t\le2t_1-t_0$ is:
\begin{align}\label{eq:q1stress2}
\tau_{yx}(t)=\left[\tau_{yx}^1+\frac{\eta_0\gamma_0a(1+a\lambda_2)}{4}\right]e^{-a(t-t_1)}-\frac{\eta_0\gamma_0a^2(1-a\lambda_2)}{2}(t-t_1)e^{-a(t-t_1)}-\frac{\eta_0\gamma_0a(1+a\lambda_2)}{4}e^{a(t-t_1)},
\end{align}
with $\tau_{yx}^1$ equal to the value of $\tau_{yx}(t_1)$ obtained from Eq. (\ref{eq:q1stress1}) to ensure continuity of the shear stress at $t=t_1$. We note that the Newtonian limit $\lambda_1=\lambda_2\rightarrow0$ [$a\rightarrow\infty$] is {\it not} defined for ${\cal G}=1$, since the above expressions diverge in the limit. It appears that the singular case ${\cal G}=1$ may be unphysical. In Fig. \ref{fig:qq}(a), we plot Eqs. (\ref{eq:q1stress1}) and (\ref{eq:q1stress2}) for the representative parameter values $a=0.5s^{-1}$, $\gamma_0=1$, $\eta_0=1\hspace{0.1cm}\textrm{Pa.s.}$, $\lambda_1=2s$, $\lambda_2=1s$ as a red curve, and compare with the numerical solution (solid blue circles). The shear stress is plotted as a function of the normalized time for $10$ cycles of oscillation in Fig. \ref{fig:qq}(b).

\begin{figure}[ht]
\includegraphics[height=2.25in]{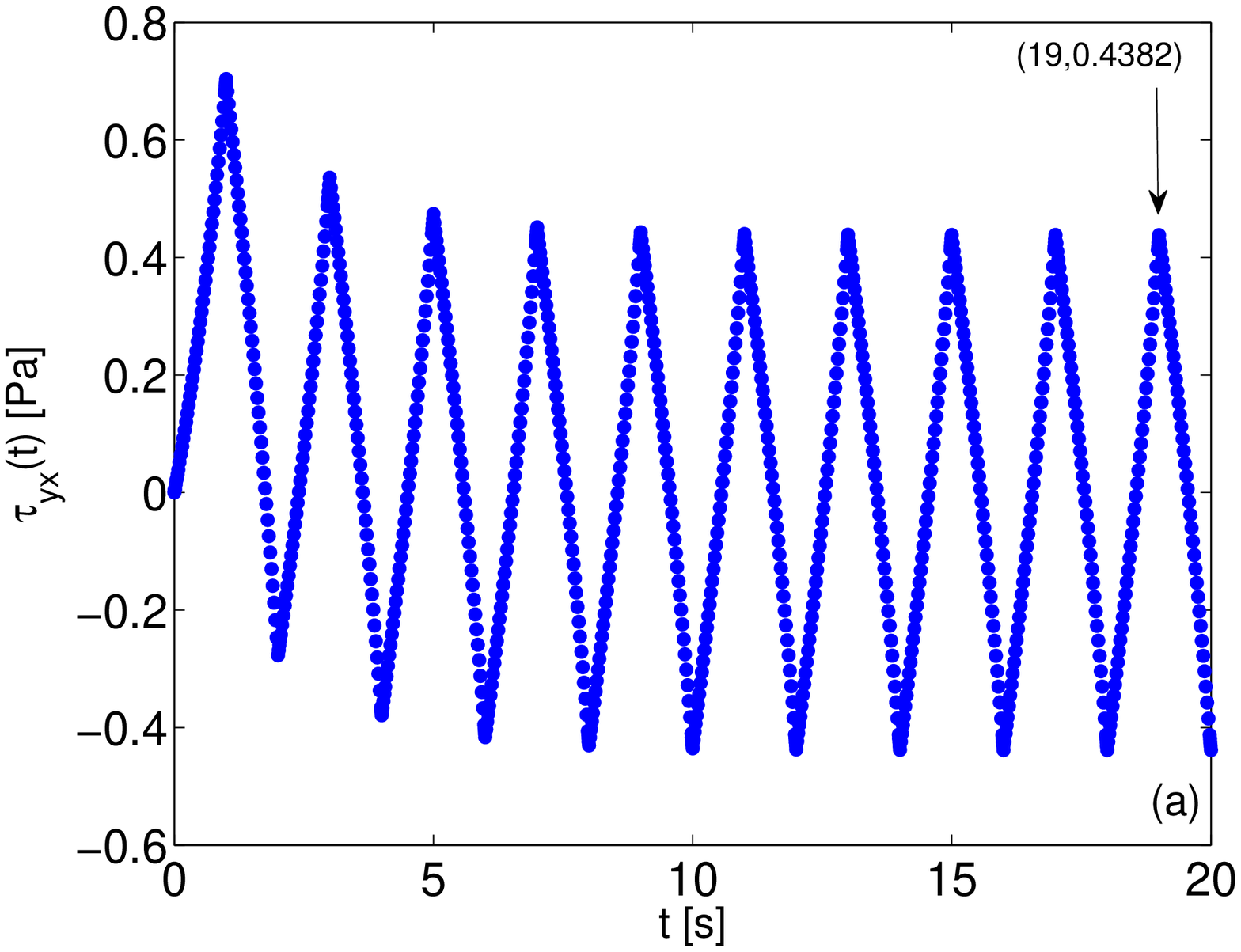}
\includegraphics[height=2.25in]{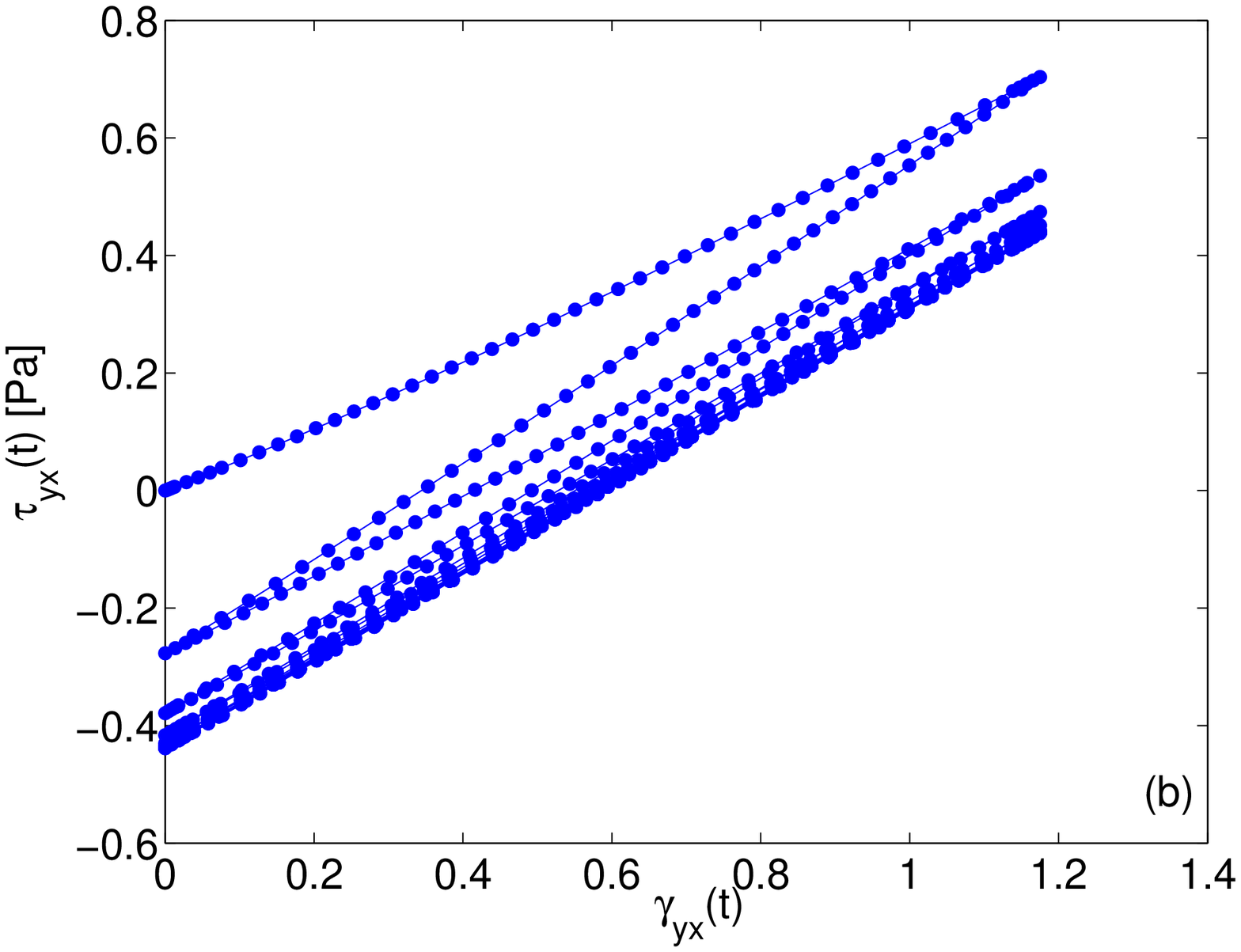}
\caption{A plot of the shear stress $\tau_{yx}(t)$ as a function of the time $t$ for 10 cycles of oscillation with $a=1s^{-1}, \gamma_0=1, \eta_0=1\hspace{0.1cm}\textrm{Pa.s.}, \lambda_1=2s, \lambda_2=1s$ (here ${\cal G}=2$). The coordinates of the peak shear stress for the tenth cycle of oscillation are indicated, for purposes of comparison with the analytically calculated value (see text for discussion). (b) A plot of the shear stress $\tau_{yx}(t)$ as a function of the shear strain ${\gamma}_{yx}(t)$ for the same parameter values as in (a).}
\label{fig:asymp}
\end{figure}

The shear stress, after an initial transient, approaches a stable oscillatory state (which has been termed {\it alternance}\cite{Giacomin}) in both Figs. \ref{fig:ss_verify}(b) and \ref{fig:qq}(b) and we now find the analytical expression for the peak shear stress at an arbitrary time $t=n$ ($n$ is a positive integer). For the case ${\cal G}\ne1$, we plot the shear stress as a function of the time for 10 cycles of oscillation in Fig. \ref{fig:asymp}(a) and refer the reader to the figure for the steps that follow : In the first cycle of oscillation [$t_0=0s$, $t_1=1s$], the numerical value of the shear stress at $t=1s$ is $\tau_{yx}(1)=(\tau_{yx}^0+\alpha)e^{-1/\lambda_1}+(\beta\mu_0+\gamma\mu_1)$, with $\alpha\equiv\frac{\eta_0\gamma_0a(1-a^2\lambda_1\lambda_2)}{(\lambda_1a)^2-1}, \beta\equiv\frac{\eta_0\gamma_0a}{(\lambda_1a)^2-1}, \gamma\equiv\frac{\lambda_2\eta_0\gamma_0a^2}{(\lambda_1a)^2-1}$, $\mu_0=\lambda_1a \sinh(a)- \cosh(a)$, $\mu_1=\lambda_1a \cosh(a)- \sinh(a)$. The numerical value of the shear stress at $t=2s$ is $\tau_{yx}(2)=(\tau_{yx}^1-\alpha)e^{-1/\lambda_1}-(\beta\mu_0+\gamma\mu_1)=(\tau_{yx}^0+\alpha)e^{-2/\lambda_1}+(\beta\mu_0+\gamma\mu_1)e^{-1/\lambda_1}-\alpha e^{-1/\lambda_1}-(\beta\mu_0+\gamma\mu_1)$, on substituting for $\tau_{yx}^1=\tau_{yx}(1)$. The procedure may now be repeated iteratively for $t=3s$ to obtain $\tau_{yx}(3)=(\tau_{yx}^0+\alpha)e^{-3/\lambda_1}+(\beta\mu_0+\gamma\mu_1)e^{-2/\lambda_1}-\alpha e^{-2/\lambda_1}-(\beta\mu_0+\gamma\mu_1)e^{-1/\lambda_1}+\alpha e^{-1/\lambda_1}+(\beta\mu_0+\gamma\mu_1)$. By inspection of the foregoing expressions, we write below the expression for the numerical value of the peak shear stress at $t=n$ ($n$ is a positive integer): 
\begin{align}
\tau_{yx}(n)=\tau_{yx}^0e^{-n/\lambda_1}+\left[\alpha+e^{1/\lambda_1}(\beta\mu_0+\gamma\mu_1)\right]\sum_{m=1}^n(-1)^{m+n}e^{-m/\lambda_1}. 
\label{eq:asymp}
\end{align}
From Eq. (\ref{eq:asymp}) we find that $\tau_{yx}(19)=0.4382$ Pa for the representative parameter values $a=1s^{-1}$, $\gamma_0=1$, $\eta_0=1\hspace{0.1cm}\textrm{Pa.s.}$, $\lambda_1=2s$, $\lambda_2=1s$. The calculated value is found to agree with the numerical value of $\tau_{yx}(19)$ in Fig. \ref{fig:asymp}.
In Fig. \ref{fig:asymp}(b), we plot the shear stress as a function of the shear strain for the same values as stated above, and note the approach to a stable oscillatory state, with a nearly elastic response at long times. For ${\cal G}=1$, using a similar iterative procedure, we find that the numerical value of the shear stress at $t=n$ ($n$ is a positive integer) is given by:
\begin{align}
\tau_{yx}(n)=(\tau_{yx}^0-\alpha)e^{-an}+\alpha e^{-a(n-1)}+\alpha(-1)^n(1-e^{a})+a\beta\sum_{m=1}^n (-1)^{m+n}e^{-am},
\end{align}
with $\alpha=\frac{\eta_0\gamma_0a(1+a\lambda_2)}{4}$, $\beta=\frac{\eta_0\gamma_0a(1-a\lambda_2)}{2}$.\\\\
\subsection{Normal stress $\tau_{xx}(t)$}
\begin{figure}[ht]
\includegraphics[height=2.25in]{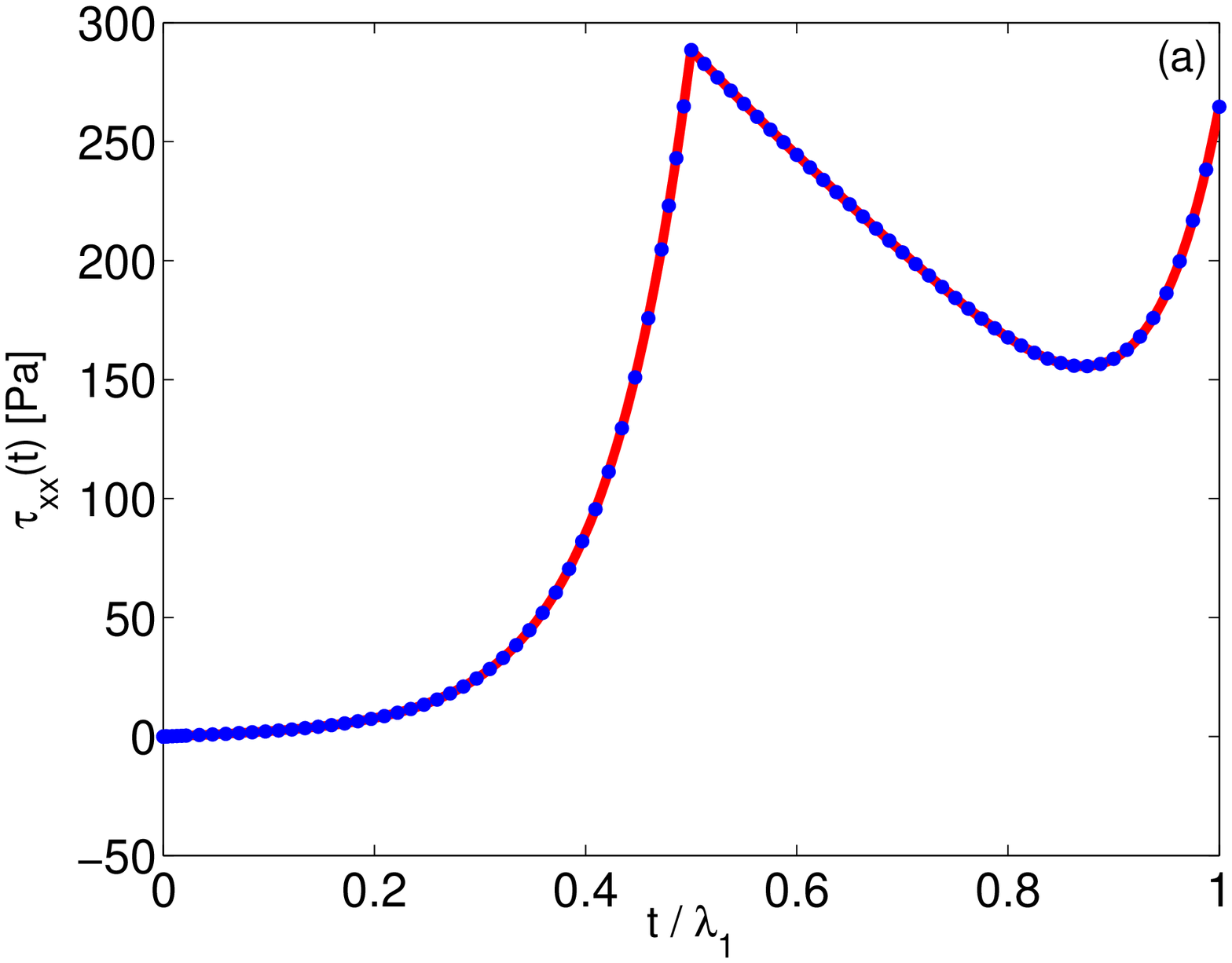}
\includegraphics[height=2.25in]{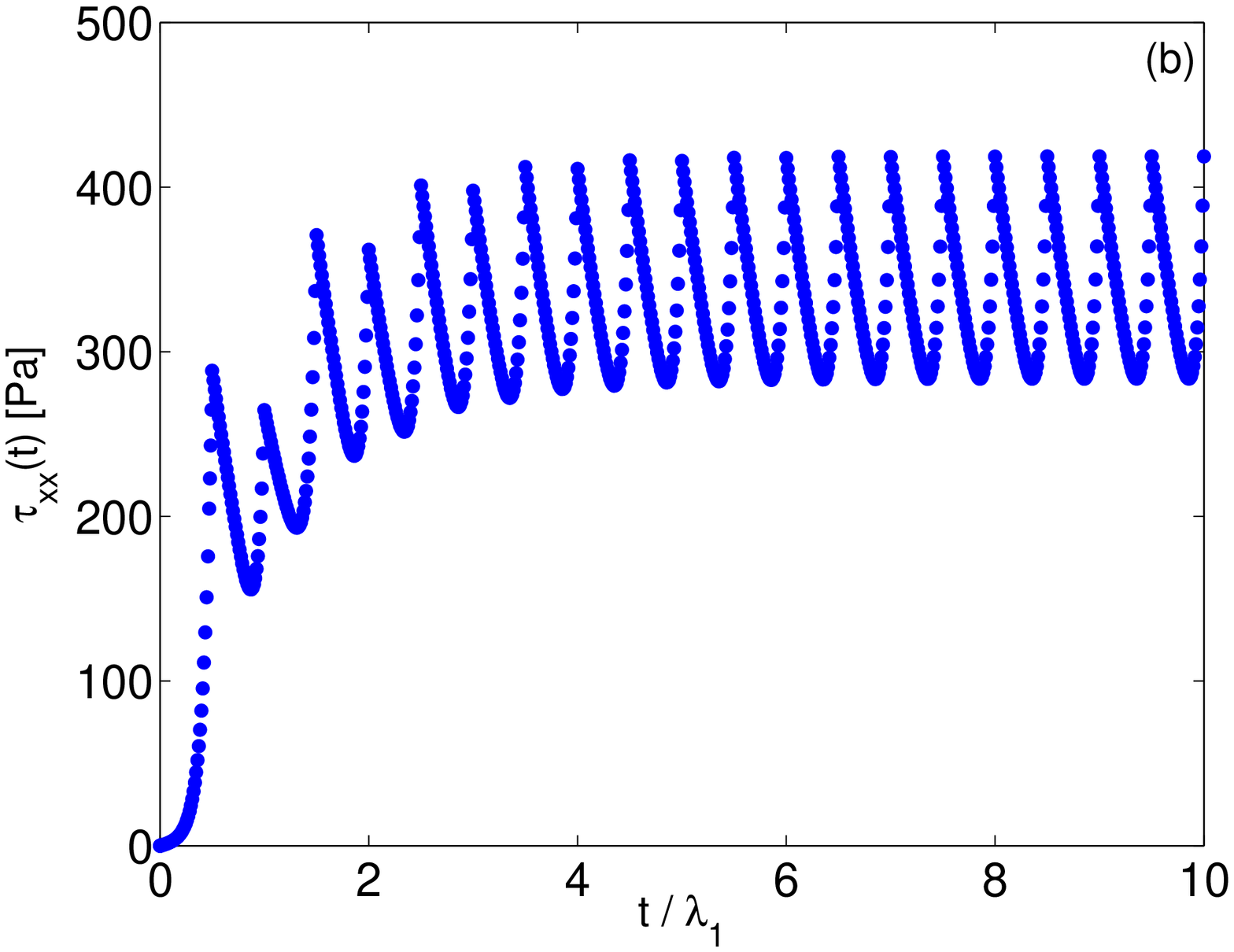}
\caption{(a) A plot of the analytical solution (red curve) for the normal stress $\tau_{xx}(t)$ as a function of the normalized time $t/\lambda_1$ along with the numerical solution (solid blue circles) of the Oldroyd-B model for $a=3s^{-1}$, $\gamma_0=1$, $\eta_0=1\hspace{0.1cm}\textrm{Pa.s.}$, $\lambda_1=2s$, 
$\lambda_2=1s$ (here ${\cal G}=6$). (b) A plot of the normal stress $\tau_{xx}(t)$ as a function of the normalized time $t/\lambda_1$ with the same parameters as above, for $10$ cycles of oscillation.}
\label{fig:qq1}
\end{figure}

The normal stress may be calculated by substituting Eqs. (\ref{eq:oldss1}) and (\ref{eq:oldss2}) into Eq. (\ref{eq:oldroydns}) and solving the resultant linear first-order ordinary differential equation. For $t_0\le t\le t_1$, ${\cal G}\ne\{0.5,1\}$ the normal stress is found to be:
\begin{align}\label{eq:ns1}
\tau_{xx}(t)=C_1e^{-t/\lambda_1}+2\gamma_0e^{-(t-t_0)/\lambda_1} \sinh(a(t-t_0))\left[\tau_{yx}^0+\frac{\eta_0\gamma_0a(1-a^2\lambda_1\lambda_2)}{(\lambda_1a)^2-1}\right]\\\nonumber
+\frac{\eta_0(\gamma_0a)^2}{4(\lambda_1a)^2-1}\left[\lambda_2+\frac{\lambda_1(a^2\lambda_1\lambda_2-1)}{(\lambda_1a)^2-1}\right][4(\lambda_1a)^2-1+2\lambda_1a \sinh(2a(t-t_0))- \cosh(2a(t-t_0))]\\\nonumber
+\frac{\eta_0(\lambda_1-\lambda_2)(\lambda_1a)(\gamma_0a)^2}{[(\lambda_1a)^2-1][(4(\lambda_1a)^2-1]}[2\lambda_1a \cosh(2a(t-t_0))- \sinh(2a(t-t_0))],
\end{align}
with $C_1=e^{t_0/\lambda_1}\left[\tau_{xx}^0+\frac{\eta_0(\gamma_0a)^2(2-4(\lambda_1a)^2)}{4(\lambda_1a)^2-1}\left(\lambda_2+\frac{\lambda_1(a^2\lambda_1\lambda_2-1)}{(\lambda_1a)^2-1}\right)+\frac{2\eta_0(\lambda_2-\lambda_1)(\lambda_1a)^2(\gamma_0a)^2}{[(\lambda_1a)^2-1][4(\lambda_1a)^2-1]}\right].$
For $t_1\le t\le 2t_1-t_0$, ${\cal G}\ne\{0.5,1\}$ the normal stress is found to be:
\begin{align}\label{eq:ns2}
\tau_{xx}(t)=C_2e^{-t/\lambda_1}-2\gamma_0e^{-(t-t_1)/\lambda_1} \sinh(a(t-t_1))\left[\tau_{yx}^1-\frac{\eta_0\gamma_0a(1-a^2\lambda_1\lambda_2)}{(\lambda_1a)^2-1}\right]\\\nonumber
+\frac{\eta_0(\gamma_0a)^2}{4(\lambda_1a)^2-1}\left[\lambda_2+\frac{\lambda_1(a^2\lambda_1\lambda_2-1)}{(\lambda_1a)^2-1}\right][4(\lambda_1a)^2-1+2\lambda_1a \sinh(2a(t-t_1))- \cosh(2a(t-t_1))]\\\nonumber
+\frac{\eta_0(\lambda_1-\lambda_2)(\lambda_1a)(\gamma_0a)^2}{[(\lambda_1a)^2-1][(4(\lambda_1a)^2-1]}
[2\lambda_1a \cosh(2a(t-t_1))- \sinh(2a(t-t_1))],
\end{align}
with $C_2=e^{t_1/\lambda_1}\left[\tau_{xx}^1+\frac{\eta_0(\gamma_0a)^2(2-4(\lambda_1a)^2)}{4(\lambda_1a)^2-1}\left(\lambda_2+\frac{\lambda_1(a^2\lambda_1\lambda_2-1)}{(\lambda_1a)^2-1}\right)+\frac{2\eta_0(\lambda_2-\lambda_1)(\lambda_1a)^2(\gamma_0a)^2}{[(\lambda_1a)^2-1][4(\lambda_1a)^2-1]}\right]$. Here $\tau_{xx}^1$ is the value of $\tau_{xx}(t_1)$ obtained from Eq. ($\ref{eq:ns1}$), which ensures continuity of the normal stress at $t=t_1$. As for the case of the shear stress, special solutions for the normal stress may be found for the singular cases 
${\cal G}=\{0.5,1\}$, but are not discussed here. In Fig. \ref{fig:qq1}(a), we plot Eqs. (\ref{eq:ns1}) and (\ref{eq:ns2}) for the representative parameter values $a=3s^{-1}$, $\gamma_0=1$, $\eta_0=1\hspace{0.1cm}\textrm{Pa.s.}$, $\lambda_1=2s$, $\lambda_2=1s$ as a red curve and compare with the numerical solution (solid blue circles) of the Oldroyd-B model. The evolution in the normal stress with the same parameter values is plotted as a function of the normalized time for $10$ cycles of oscillation in Fig. \ref{fig:qq1}(b). The normal stress is shown to approach a stable oscillatory state after a complex initial transient response and the analytical expression for the numerical value of the peak normal stress may also be found via an iterative procedure, as outlined above for the shear stress.

We note that for the Newtonian limit of the Oldroyd-B model: $\lambda_1=\lambda_2\rightarrow0$, Eqs. (\ref{eq:oldss1}), (\ref{eq:oldss2}) reduce to $\tau_{yx}(t)=\eta_0\gamma_0a \cosh(a(t-t_0))=\eta_0\dot{\gamma}_{yx}(t)$, while Eqs. (\ref{eq:ns1}), (\ref{eq:ns2}) reduce to $\tau_{xx}=0$, which are consistent with the anticipated expressions\cite{Bird} for a Newtonian fluid. 
\subsection{Derived quantities}
In oscillatory shear, the kinetic energy dissipated per unit volume per cycle of imposed strain oscillation is given by $\epsilon\equiv\int\tau_{yx}d\gamma_{yx}$\cite{Bird}. For PES flows, we use Eqs. (\ref{eq:oldss1}) and (\ref{eq:oldss2}) to calculate the dissipation rate for ${\cal G}\ne1, t_0\le t\le 2t_1-t_0$:
\begin{align}
\epsilon=\frac{\gamma_0\lambda_1a}{(\lambda_1a)^2-1}\left[(\tau_{yx}^0-\tau_{yx}^1)+\frac{2\eta_0\gamma_0a(1-a^2\lambda_1\lambda_2)}{(\lambda_1a)^2-1}\right]\left[e^{-(t_1-t_0)/\lambda_1}(\lambda_1a \sinh(a(t_1-t_0))+ \cosh(a(t_1-t_0)))-1\right]\\\nonumber
+\frac{\eta_0(\gamma_0a)^2(\lambda_1-\lambda_2)}{2[(\lambda_1a)^2-1]}\left[\cosh(2a(t_1-t_0))-1\right]+\frac{\eta_0\gamma_0^2a(a^2\lambda_1\lambda_2-1)}{2[(\lambda_1a)^2-1]}\left[2a(t_1-t_0)+ \sinh(2a(t_1-t_0))\right].
\end{align}
In the Newtonian limit $\lambda_1=\lambda_2\rightarrow0$, the above expression reduces to $\epsilon=\frac{\eta_0\gamma_0^2a}{2}[2a(t_1-t_0)+ \sinh(2a(t_1-t_0))]$. 

\begin{figure}[ht]
\includegraphics[height=2.25in]{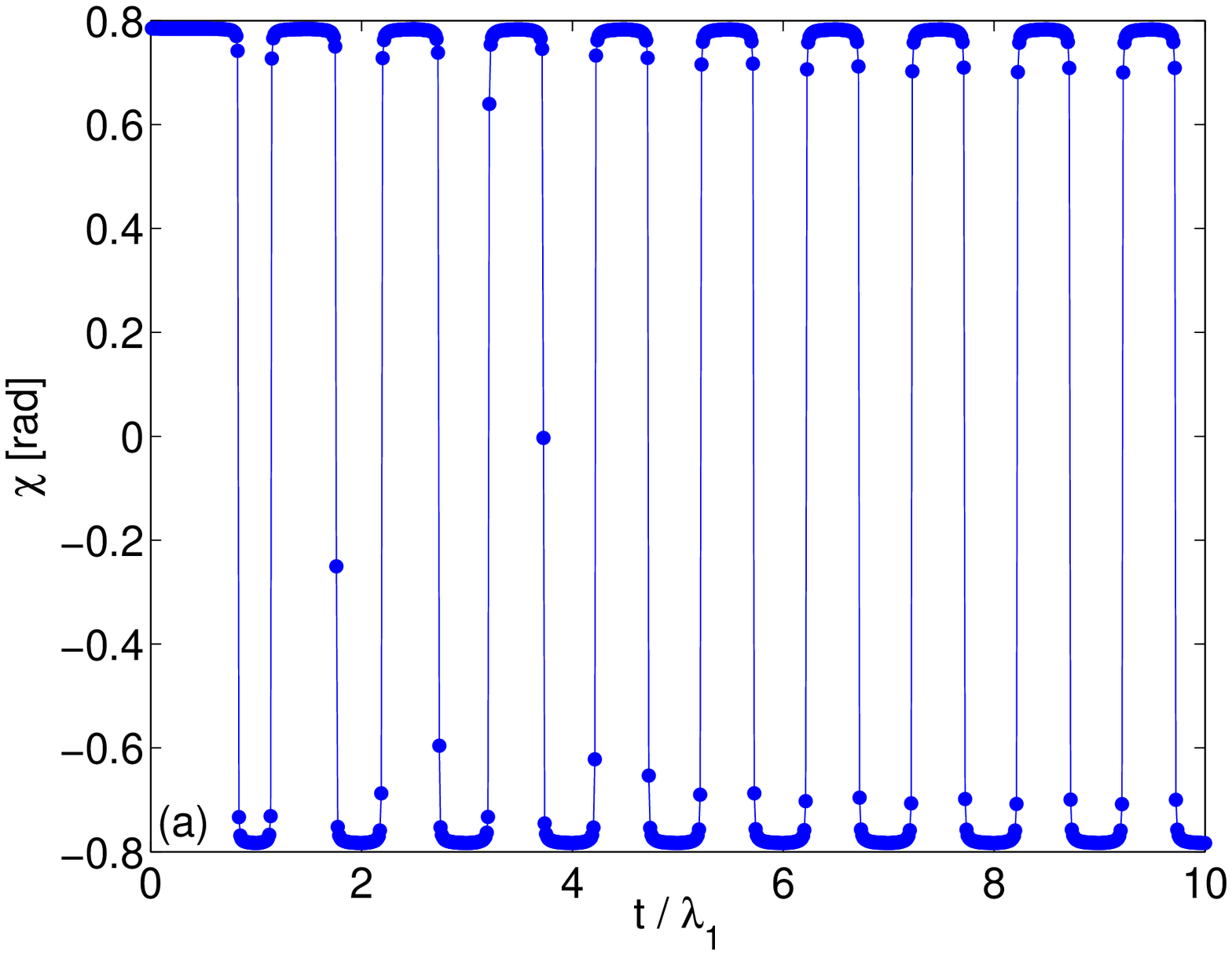}
\includegraphics[height=2.25in]{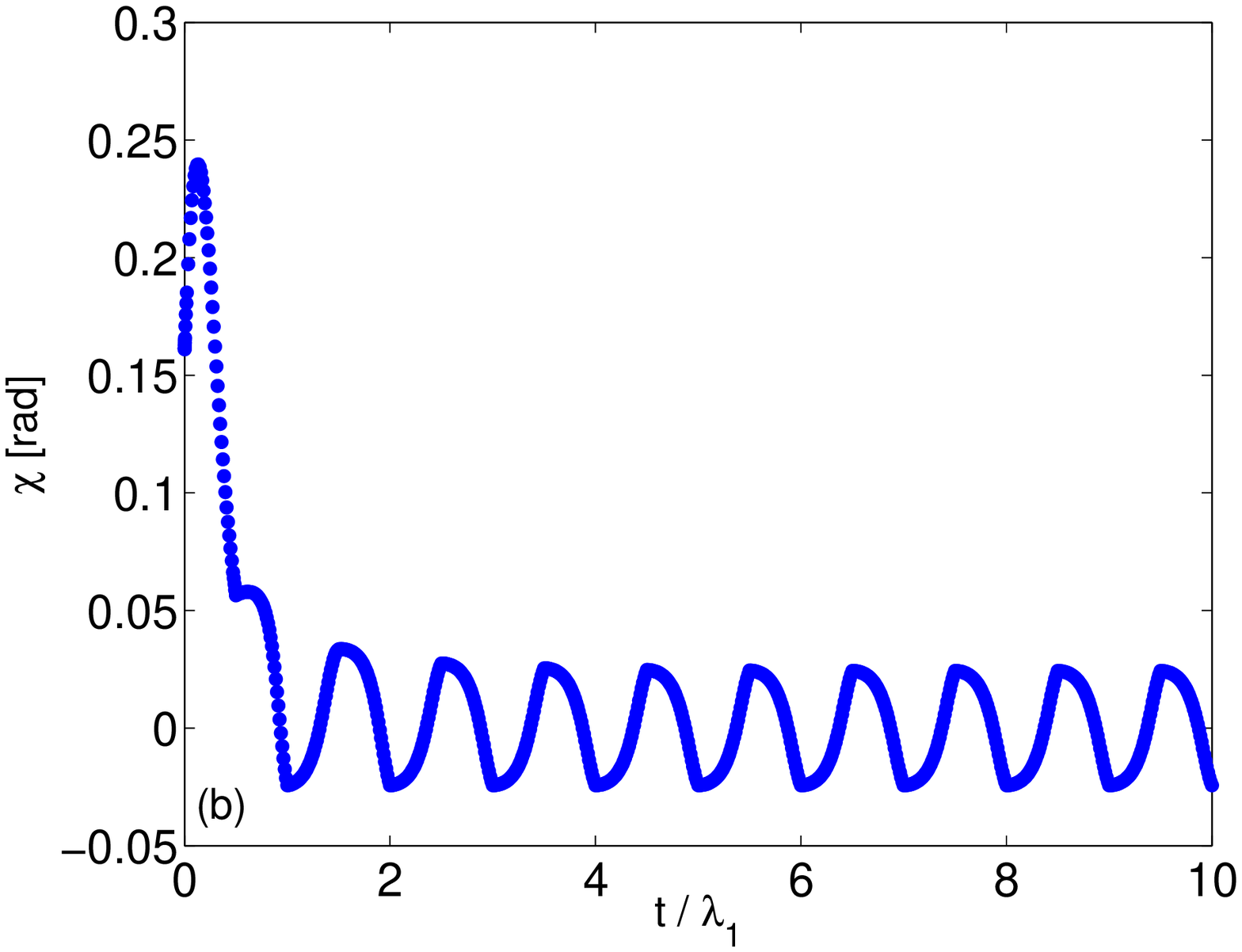}
\caption{(a) A plot of the extinction angle $\chi(t)\equiv(0.5)tan^{-1}[2\tau_{yx}(t)/\tau_{xx}(t)]$ as a function of the normalized time $t/\lambda_1$ for $10$ cycles of oscillation of PES flow with the parameter values $a=0.001s^{-1}, \gamma_0=1, \eta_0=1\hspace{0.1cm}\textrm{Pa.s.}, \lambda_1=2s, \lambda_2=1s$ (here ${\cal G}=0.002$). (b) A plot of the extinction angle as a function of the normalized time for $10$ cycles of oscillation of PES flow with the parameter values $a=3s^{-1}, \gamma_0=1, \eta_0=1\hspace{0.1cm}\textrm{Pa.s.}, \lambda_1=2s, \lambda_2=1s$ (here ${\cal G}=6$).}
\label{fig:extangle}
\end{figure}

A measure of the average orientation of polymer chains within a viscoelastic fluid flow is given by the extinction angle $\chi(t)\equiv\frac{1}{2}tan^{-1}[2\tau_{yx}(t)/N_1(t)]$ \cite{Neergaard}. In Fig. \ref{fig:extangle}(a), we plot $\chi(t)$ as a function of the normalized time for the parameter values $a=0.001s^{-1}$, $\gamma_0=1$, $\eta_0=1\hspace{0.1cm}\textrm{Pa.s.}$, $\lambda_1=2s$, $\lambda_2=1s$. The limit $a\rightarrow0$ is found to be equivalent to the Newtonian limit of the Oldroyd-B model, and the graph of the extinction angle alternates between $\chi\approx\pi/4$ and $\chi\approx-\pi/4$ every half-cycle. In Fig. \ref{fig:extangle}(b), we plot the extinction angle as a function of the normalized time for similar parameter values as chosen for Figs. \ref{fig:ss_verify}(b) and \ref{fig:qq1}(b). We find that after an initial transient, the polymer chains are stretched and rotated periodically about the flow direction as a result of the vorticity in the flow. 
\subsection{Material function for PES flow}
In earlier studies\cite{Zulle, Dealy,Jobling,Lodge} several different material functions have been proposed for exponential shear. Z\"ulle {\it et al.}\cite{Zulle} used a material function $\eta_Z$ (where the subscript $Z$ indicates ``Z\"ulle") which varies with the choice of the coordinate system employed and does not account for contributions from the normal stresses:
\begin{eqnarray}\label{eq:zz}
\eta_{Z}(t)\equiv\frac{\tau_{yx}(t)}{a}.
\end{eqnarray}
Although $\eta_Z$ shows shear-thickening behavior for the exponential shear of viscoelastic fluids, Samurkas {\it et al.}\cite{Larson} pointed out that the material function also shear-thickens for a Newtonian (purely viscous) fluid, on account of the exponential growth of the shear stress, so unbounded shear stress growth does not imply shear-thickening behavior of the fluid just the exponential increase of the deformation rate. For {\it steady} shear flows, Jobling and Roberts\cite{Jobling}, and independently, Lodge and Meissner\cite{Lodge} proposed a shear viscosity $\eta_{LM}$ (subscript $LM$ indicates ``Lodge-Meissner") which is independent of the choice of any special coordinate system used to define the flow, and is given by the ratio of the difference of the first two components of the principal stress tensor divided by the difference of the first two components of the principal strain-rate tensor, which may be rewritten in terms of the first-normal stress difference and the shear stress:
\begin{eqnarray}\label{eq:lmvisc}
\eta_{LM}\equiv\frac{\sqrt{N_1^2+4\tau_{yx}^2}}{2|\dot{\gamma}_{yx}|}.
\end{eqnarray}
Doshi and Dealy\cite{Dealy} suggested that Lodge and Meissner's definition of the shear viscosity be applied to the case of {\it unsteady} shear flow, with the difference of the (time-dependent) principal strain-rates in the denominator:
\begin{eqnarray}\label{eq:dd}
\eta_{DD}(t)\equiv\frac{\sqrt{N_1^2(t)+4\tau_{yx}^2(t)}}{2|\dot{\gamma}_{yx}(t)|},
\end{eqnarray}
where the subscript $DD$ indicates ``Doshi-Dealy". 

\begin{figure}[ht]
\includegraphics[height=2.25in]{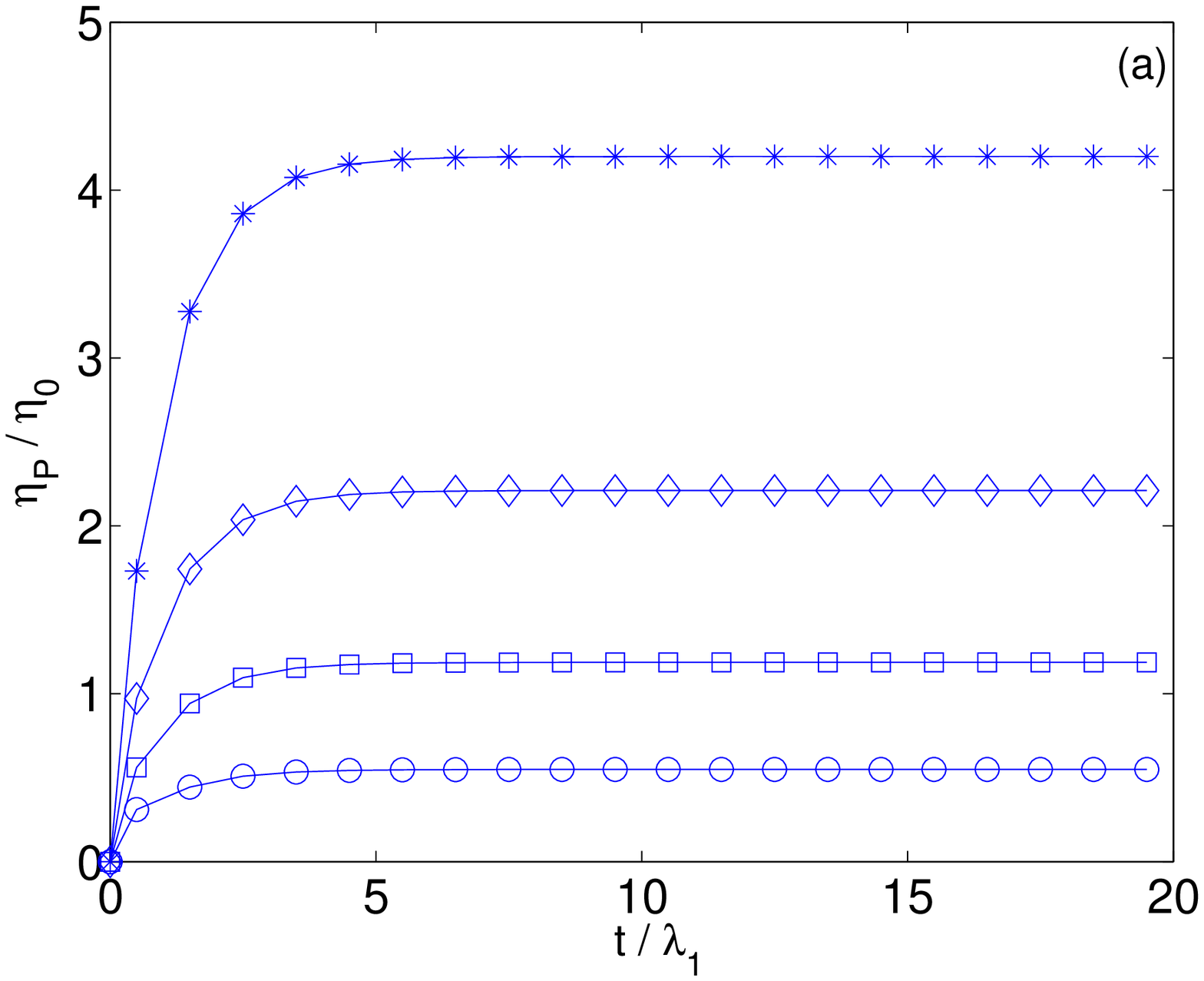}
\includegraphics[height=2.25in]{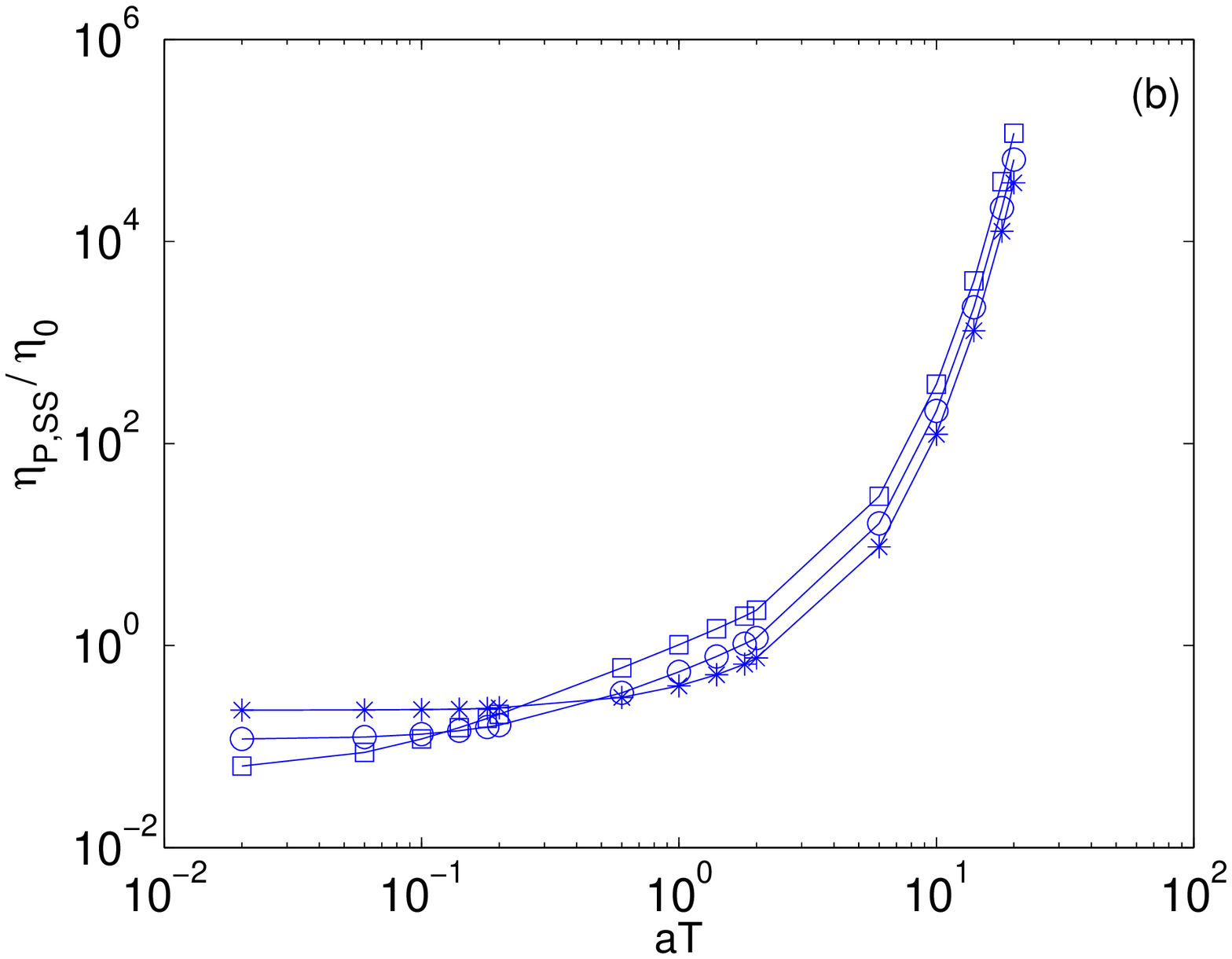}
\caption{(a) A plot of the normalized PES viscosity $\eta_{P}/\eta_0$ as a function of the normalized time $t/\lambda_1$ for 20 cycles of oscillation with $\gamma_0=1, \eta_0=1\hspace{0.1cm}\textrm{Pa.s}, \lambda_1=2s, \lambda_2=1s$ and varying exponential growth rate $a$ with the legend: circle ($a=0.5s^{-1}$), square ($a=1s^{-1}$), diamond ($a=1.5s^{-1}$), star ($a=2s^{-1}$). (b) A plot of the steady-state normalized PES viscosity $\eta_{P,SS}/\eta_0$ as a function of the normalized exponential growth rate $aT$ (here $T=2(t_1-t_0)$ is the time-period of one flow-cycle) for varying relaxation times $\lambda_1, \lambda_2\equiv\lambda_{1}/2$ with the legend: star ($\lambda_1=1s$), circle ($\lambda_1=2s$), square ($\lambda_1=4s$) and $\gamma_0=1, \eta_0=1\hspace{0.1cm}\textrm{Pa.s.}$}
\label{fig:compvisc}
\end{figure}

We note that Eqs. (\ref{eq:zz}), (\ref{eq:lmvisc}), and (\ref{eq:dd}) are clearly inappropriate for the case of PES flow, which is a {\it periodic unsteady} shear flow with a discontinuous strain-rate at the half-period time $t_1$. We propose a frame-invariant material function, based on the Lodge-Meissner viscosity, which we call the Generalized Lodge-Meissner viscosity $\eta_{GLM}$, suitable for any periodic unsteady shear flow, calculated over an oscillation cycle:
\begin{eqnarray}\label{eq:glm}
\eta_{GLM}\equiv\frac{\oint\sqrt{N_1^2(t)+4\tau_{yx}^2(t)}\hspace{0.1cm}dt}{2\oint|\dot{\gamma}_{yx}(t)|\hspace{0.1cm}dt}.
\end{eqnarray}
The Generalized Lodge-Meissner viscosity is calculated by integrating the difference of the first two components of the principal stress tensor over an oscillation cycle and dividing by the integrated difference of the first two components of the principal strain-rate tensor over the same cycle. We note that for a Newtonian fluid (defined by $\tau_{yx}(t)\equiv\eta\dot{\gamma}_{yx}(t)$ and $N_1(t)=0$), the Generalized Lodge-Meissner viscosity reduces to the anticipated result 
$\eta_{GLM}=\eta$. For PES flow, upon substituting for the strain-rate in the denominator of Eq. (\ref{eq:glm}), we obtain the PES viscosity $\eta_P$ (where the subscript $P$ indicates ``PES flow") calculated over a single oscillation cycle:
\begin{eqnarray}\label{eq:PES}
\eta_{P}\equiv\frac{\int_{t_0}^{2t_1-t_0}\sqrt{N_1^2(t)+4\tau_{yx}^2(t)}\hspace{0.1cm}dt}{4\gamma_0 \sinh[a(t_1-t_0)]}.
\end{eqnarray}
We plot Eq. (\ref{eq:PES}) with the representative parameter values $\gamma_0=1, \eta_0=1\hspace{0.1cm}\textrm{Pa.s.}, \lambda_1=2s, \lambda_2=1s$ and varying exponential growth rates $a$ in Fig. \ref{fig:compvisc}(a). For convenience, we plot the calculated value of $\eta_{P}$ over each cycle, at the mid-point of the cycle. We note that the PES viscosity always approaches a steady-state value and is continuous. In Fig. \ref{fig:compvisc}(b), we plot the calculated steady-state value  (after $100$ cycles of oscillation) of the normalized PES viscosity $\eta_{P,SS}/\eta_0$ as a function of the normalized exponential growth rate $aT$ (here $T=2(t_1-t_0)$ is the time-period) for $\gamma_0=1, \eta_0=1\hspace{0.1cm}\textrm{Pa.s.}$ and varying relaxation times $\lambda_1, \lambda_2\equiv\lambda_{1}/2$. We note that the steady-state PES viscosity grows faster than the linear-viscoelastic prediction (which corresponds to $\eta_{P,SS}/\eta_0=1$), indicating shear-thickening behavior of the viscoelastic fluid. By contrast, Neergaard {\it et al.}\cite{Neergaard} concluded from numerical studies with a reptation model that (aperiodic) exponential shear does not exhibit shear-thickening ``in any practicable experiment for linear, entangled polymer chains in shear flow".

\subsection{Experimental results}
Our experiments were carried out at room temperature on a strain-controlled Advanced Rheometric Expansion System rheometer (ARES-LS$818409$, TA Instruments, United States). We used a cone-plate assembly with a cone diameter of $50$mm (cone angle=$0.04$rad). The rheometer permits direct acquisition of dc voltage signals from the torque transducer, the optical encoder, and the transducer normal force sensor through BNC connectors in the rear panel of the instrument. These unprocessed voltage signals are not noise-filtered or corrected for inertia and compliance of the torque transducer. Data was acquired at 16-bit resolution through an analog-to-digital card (NI PCI-6032E, National Instruments, United States) coupled with a Labview (National Instruments, United States) code at a sampling rate  of $10^3$ points per second. The acquired signal was filtered for noise using a Savitzky-Golay filter and calibrated to find formulae which were used to convert the voltage values to quantities of physical interest. The calibration curves used were $y=0.035x$ ($x$ in volts, $y$ in Newton-metre) for the torque, $y=0.1x$ ($x$ in volts, $y$ in radians) for the deflection angle and $y=-4.05x$ ($x$ in volts, $y$ in Newtons) for the normal force. The values of the stress (in Pascals), the strain and the first-normal stress difference (in Pascals) were calculated from the torque, the deflection angle, and the normal force respectively, using conversion factors appropriate to the measuring system geometry and torque transducer employed. The rheometer can be easily programmed to accomplish exponential shearing flow by supplying the desired strain profile to the proprietary software used to control the instrument. However, due to a hardware limitation of the rheometer, continuous oscillation cycles of PES flow could not be achieved: The rheometer accomplishes two complete oscillation cycles of PES flow, after which there is a delay of approximately $25s$ (during which the specified waveform is sent to the instrument's memory) before the next two cycles commence; which is why we show only two continuous cycles in the plot that follows.

\begin{figure}[ht]
\includegraphics[height=2.25in]{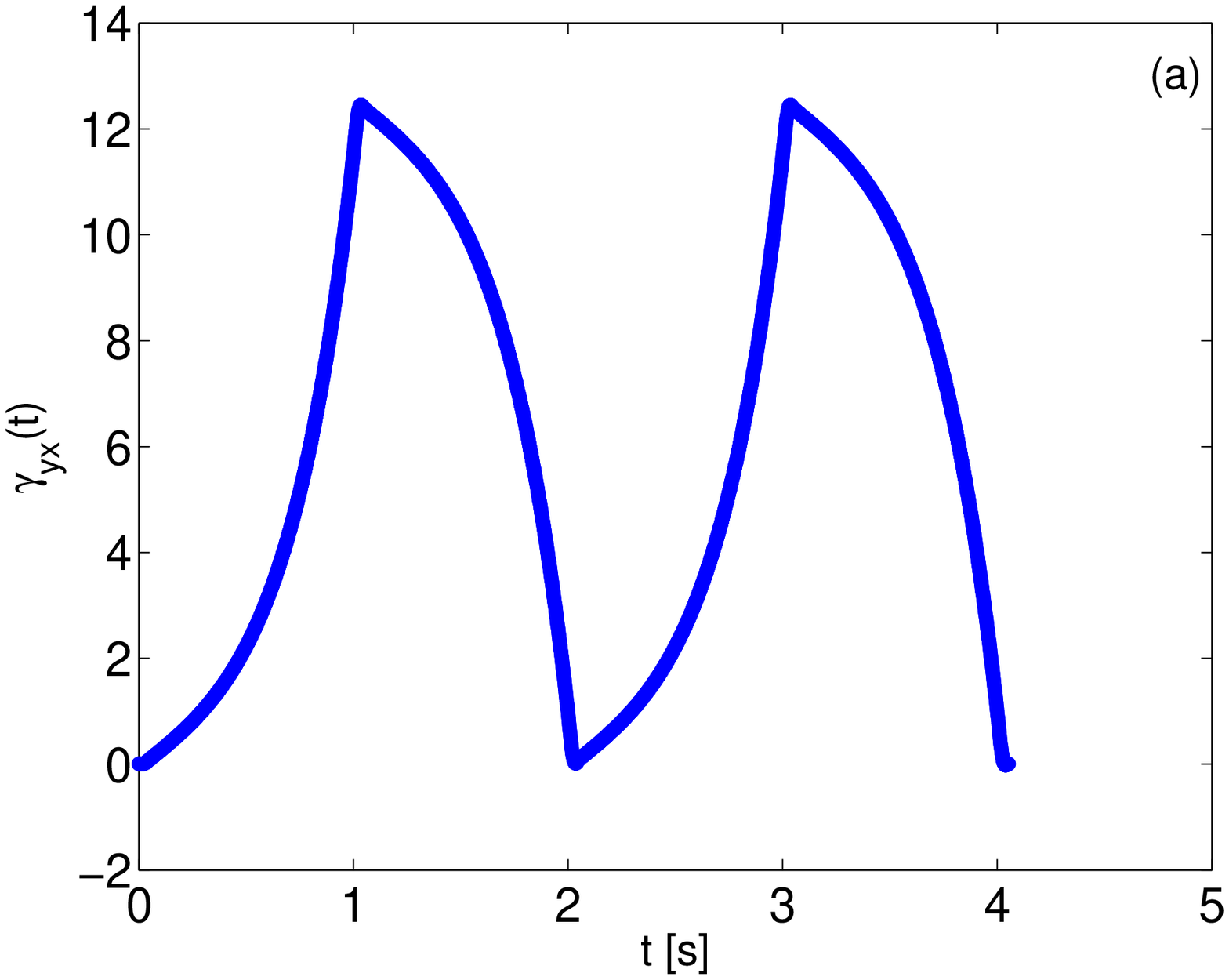}
\includegraphics[height=2.25in]{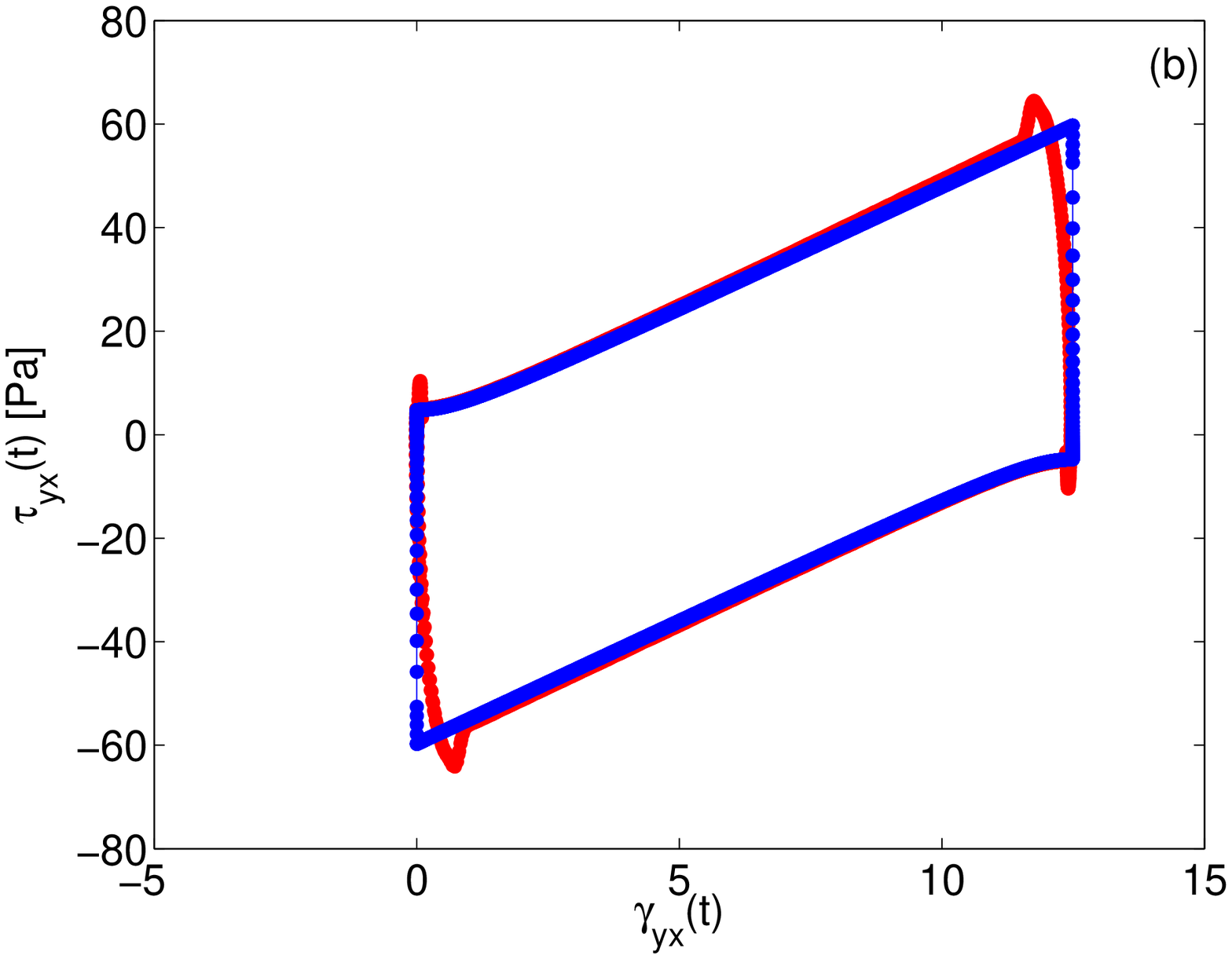}
\caption{(a) A plot of the imposed shear strain $\gamma_{yx}(t)$ as a function of the time $t$ for two cycles of oscillation of PES flow with $a=3.2s^{-1}, \gamma_0=1$. (b) A comparison of the response shear stress $\tau_{yx}(t)$ for two cycles of oscillation as a function of the shear strain $\gamma_{yx}(t)$ for the Newtonian mineral oil N$1000$ (red curve) with the numerically calculated shear stress (blue curve) having model parameter values: $a=3.22s^{-1}, \eta_0=1.48$ Pa.s., 
$\gamma_0=1, \lambda_1=\lambda_2=10^{-4}$, which approximate the Newtonian limit.} 
\label{fig:n1000}
\end{figure}

For our tests we used a viscous Newtonian mineral oil N1000 (Cannon Instrument Company, United States) with the stated zero-shear-rate viscosity $\eta_0=2.01$ Pa.s at $25^\circ C$.
In Fig. \ref{fig:n1000}(a), we plot two cycles of oscillation of the imposed strain as a function of the time for $\gamma_0=1$, $a=3.2s^{-1}$. The measured shear stress response for N$1000$ oil is plotted as a function of the shear strain (red curve) in Fig. \ref{fig:n1000}(b). The Lissajous plot shows that the rheometer has a lag time of approximately $0.05s$ at startup of motion of the plate. For purposes of comparison, the shear stress calculated numerically from the Oldroyd-B model with parameter values $a=3.22s^{-1}$, $\eta_0=1.48$ Pa.s., $\gamma_0=1, \lambda_1=\lambda_2=10^{-4}$ (which approximate the Newtonian limit of the model) is also shown (blue curve). The measured normal force for the Newtonian N$1000$ oil was found to be a small noisy signal, and is not shown here. 
\section{Conclusion}
To summarize, we have introduced a new rheological test protocol consisting of a periodic, exponential shear flow, and discussed the response of a viscoelastic fluid such as a dilute polymer solution that is well-described by the Oldroyd-B constitutive equation. Exact analytic expressions for the shear and the normal stress have been calculated and compared with numerical computation of the model. An appropriate material function for this periodic, oscillating flow has been defined and shown to exhibit deformation-rate hardening.\\
It is known\cite{Handbook} that fluid invasion into the well-bore of an oil rig weakens the rock formation, and can cause a collapse. Improved well-bore strengthening may be achieved by the use of drilling fluids which shear-thicken while flowing through the drill bit. The observation (or lack thereof) of shear-thickening in the drilling fluid could potentially be related to the control of fluid invasion near the well-bore. The PES flow kinematics defined in this paper may serve as a basis for distinguishing between different formulations of drilling fluids, by studying their rheological response to a well-characterized large-strain oscillating flow. We hope that our comprehensive study of a periodic flow which incorporates both rotation and rapid stretching of the fluid elements is found to be of utility within laboratory and industrial settings.
\section{Acknowledgments}
We thank Tejas Kalelkar, Bavand Keshavarz and Simon Haward for discussions, Allen Glasman of TA Instruments (United States) for technical support, Jason Maxey of Halliburton (United States) for motivational discussions.





\bibliographystyle{model1-num-names}

\end{document}